# Using Socio-economic Indicators, Smart Transit Systems, and Urban Simulator to Accelerate ZEV Adoption and Reduce VMT

MULHAM FAWAKHERJI, BRUCE RACE
*North Carolina Agricultural and Technical University, NC, United States*
*mfawakherji@ncat.edu*
*FAIA, FAICP, University of Houston, TX, United States*
*barace@Central.UH.EDU*

AND

DRISS BENHADDOU
*Alfaisal University, Riyadh, Saudi Arabia*
*dbenhaddou@alfaisal.edu*

**Abstract**. Globally, on-road transportation is the source of 15% of greenhouse gas (GHG) emissions and the cause of an estimated 385,000 premature deaths from PM2.5. The action by cities to reduce on-road emissions is critically important to meet IPCC targets, as cities are the source of 75% of global energy-related GHG emissions. In the City of Houston, Texas, on-road transportation is 48% of baseline GHG in their Climate Action Plan (CAP). To meet a 2050 target of net zero, the City Climate Action Plan (CAP) aims to reduce emissions 70% below a 2014 baseline and use renewable sources for a 30% offset. This is a challenging target, as Houston is a low density and auto-dependent city with 89% of on-road GHG emissions are from cars and small trucks and there is a low modal split for public transit. It also has uneven pattern of socio-economic indicators indicating barriers to ZEV adoption. Strategies include accelerating access to Zero Emissions Vehicles (ZEVs) and reducing Vehicles Miles Traveled (VMT) by 20% through public transit enhancements and city design. This paper shares methods for creating an on-road emissions baseline and evaluating policy strategies that use socio-economic indicators and Intelligent Transportation Systems (ITS) to accelerate adoption of ZEVs and reduce VMT. Improving the modal split is facilitated by smart parking and transit strategies that build in incentives, data management and security, and ZEV fleet vehicle and driver management. The smart transit model demonstrates potential for increased reliability, ridership, and economic performance. Policy options are discussed and evaluated, and potential actions identified. To support the evaluation of these strategies, a simulation environment was developed using the Unity 3D game engine. This virtual platform enables dynamic modeling of urban mobility conditions, allowing for the analysis and visualization of transportation behaviors and policy scenarios in a controlled, interactive setting. Auto-dependent cities striving to meet 2050 emission targets can benefit from indicators, metrics, methods, and technologies discussed.

## 1. Introduction

Transportation is one of the largest contributors to climate change and public health risks. Even though different data has been reported on the impact of on-road transport on the human health, the most conservative data show that on-road transportation contributes to 53,000 death due to PM2.5 (Li et al., 2021). In addition, it is estimated to contribute about 15% of greenhouse gas (GHG) emissions (icct, 2019). Within cities, on-road transportation accounts for 75% of this 15% share of global GHG emissions and therefore cities bear a great responsibility of reducing GHG to meet the Intergovernmental Panel on Climate Change (IPCC). Two main approaches can be taken by city managers to reduce GHG emissions, first is to promote the use of zero-Emission Vehicles (ZEVs) by citizens, second is to reduce the Vehicle Miles Traveled (VMT) through investing in public transit, land-use, and demand management (Sims et al., 2014). Achieving these goals will combine technological, socio-economical and urban planning approaches.

Many government programs promote the adoption of ZEV vehicle in the United States (Jenn, et. Al. 2018), Europe (Martins, et al., 2024), and around the world (Sheldon et al. 2024). For example, California's Battery Electric Vehicle (BEV) program was one of the firs programs in the United States to promote EV adoption. The study done by Jenn showed that an increase in $1,000 rebate corresponded to an increase of 2.6% in EV sales in the US. Even though the impact of financial incentive policies is not uniform in European countries, Martins' study shows that incentives can contribute to EV adoption. The study done by Sheldon on 23 countries around the world using data from 2010 to 2019, showed that subsidies are effectively boosting EV market penetration.

Parallel research on transport demand management highlights the value of time travelled, and strategies to reducing Vehicles Miles Travelled (VMT) through congestion pricing, parking restrictions, and transit-oriented development (Börjesson et al., 2014, Ding, C). The rapid deployment of Intelligent Transportation Systems (ITS) has added another dimension to this landscape (Wang, et al., 2024). Advances in smart parking, integrated fare collection, and real-time fleet management have been shown to improve transit reliability, optimize traffic flow, and reduce unnecessary miles traveled by vehicle.

Despite these advances, policymakers and infrastructure developers need tools that will enable them to fully understand and act upon the complex relationship between transportation systems and GHG emissions. Existing tools and modeling techniques require experts to run them and lack the ability to visualize and analyze data for non-technical users. The result is that city managers receive aggregated statistics or a fragmented knowledge that does not provide clear and granular information.

To equitably achieve 2050 greenhouse gas reduction targets from on-road transportation, cities will need to deploy strategies that utilize emerging technologies, switch fuels, and reduce vehicle miles traveled. This paper integrates planning and engineering research to demonstrate how large sprawling cities like Houston can meet aggressive climate action goals. Research includes development of socio-economic indicators for EV adoption, smart transit systems, and preparation of a baseline Urban Simulator to Accelerate ZEV Adoption and Reduce VMT.

## 2. Cities Meeting Transportation GHG and EV Adoption Targets

The action by cities to reduce on-road emissions is critically important to meet IPCC targets, as cities are the source of 75% of global energy-related GHG emissions (UN-Habitat, 2025). In the City of Houston, Texas, on-road transportation is 48% of baseline GHG in their Climate Action Plan. To meet a 2050 target of net zero, the Houston Climate Action Plan (H-CAP) aims to reduce on-road transportation emissions 70% below a 2014 baseline and use renewable sources for a 30% offset (City of Houston, 2020).

### 2.1 CITY GHG EMISSIONS AND PROTOCOLS

The H-CAP utilized the C40 protocols for collecting and measuring GHG emissions data. The Global Protocol for Community-Scale Greenhouse Gas Emission Inventories (GPC). The GPC was created by C40, ICLEI, and the World Resources Institute (WRI). It provides a consistent framework for cities to create emissions baselines, set goals, and monitor progress (Doust, 2018).

### 2.2 H-CAP BASELINE EMISSIONS INVENTORY

The H-CAP GHG baseline for 2014 includes 49% from stationary energy sources (16,454,686 MTCO2e), 48% from transportation (16,140,987 MTCO2e), and 2% from waste (818,344 MTCO2e). On-road transportation is responsible for 99% of transportation emissions (15,932,882 MTCO2e) and the balance is from off-road and rail sources. Transportation emissions were estimated using the Environmental Protection Agency's (EPA) MOtor Vehicle Emission Simulator (MOVES) model with vehicle miles traveled (VMT) provide by Texas Department of Transportation (TxDOT).

### 2.3 ON-ROAD GHG EMISSIONS 2050 TARGET

The H-CAP provides a roadmap for meeting a net zero 2050 target. It establishes 2030 (33%), 2040 (58%), and 2050 (70%) below baseline GHG reduction goals and parallel development of renewable energy offsets. It includes performance assumptions for vehicle class and VMT reduction. Goals are to be met implementing a set of strategies and related actions.

### 2.4 PUBLIC HEALTH CO-BENEFITS

On-road transportation is a primary source of PM2.5. Reduction in PM2.5 is a co-benefit of meeting GHG emission targets. PM2.5 causes a variety of health impacts to the pulmonary, cardiac, vascular, and neurological systems (Li & Managi, 2021). The National Institutes of Health estimates there were over 8,400 premature deaths in Texas due to PM2.5 (Bryan & Landrigan, 2023).

## 3. Houston the Low-Density City

Houston's GHG 2050 targets for transportation are challenging. It is a low density and auto-dependent city with 89% of on-road GHG emissions from cars and small trucks and a low modal split for public transit. It also has an uneven pattern of socio-economic indicators representing barriers to ZEV adoption (Race, Houston-CAP On-Road Transportation GHG Emissions: Baseline, BAU, and Policy Scenario Analysis, 2019).

3.1 AUTO DEPENDENCY

Houston is a fast growing and low-density city. It is expected to grow from 2.52 million in 2020 to over 3.3 million people by 2050. Houston is a global outlier region with a low density, high per capita VMT and transportation energy consumption (Rodrigue, 2024). Houston's population per square mile is approximately 3,800 compared to Los Angles' 8,500, Chicago's 12,000, and NYC's 28,000 (Maciag, 2013). Houston's low-density results in congestion, long commutes, and a large percentage of GHG emissions from personal vehicles parked in peoples' driveways. Therefore, the pathway to zero emissions from on-road transportation will require a focus on accelerating adoption of ZEV and VMT reductions.

3.2 SOCIO-ECONOMIC PATTERNS AND THE AFFORDABILITY GAP

Houston has an uneven pattern of households that can afford new more expensive electric vehicles and related benefits. This results in an affordability gap and prevents poorer households from benefiting from lower transportation costs of owning an EV.

To better understand development of policies, strategies and related actions that can accelerate equitable adoption of EVs, researchers at University of Houston prepared an EV equity index (Race, Equitable Access to EV-Mobility, 2022). The index is comprised of internal and external indicators. Internal indicators include educational attainment, poverty level households, renter households, households with less than two cars. External indicators include public charging access, cost of EV, and existing EV incentives. Using census data, researchers mapped EV Equity Index (Figure 1). This map can be used to formulate policies and strategies to accelerate adoption, including public access charging, apartment charging for renters, and financial incentives. In addition, Houston is comprised of communities that have a high percentage of households that cannot afford car loans equaling 10% of their household income. In 2022 only 19% of Harris County census tracts had median income households that could afford a new EVs. Even with Federal incentives of $4,000 for used EVs, only 44% of census tracts had median income households can afford a used EV (Figure 2).

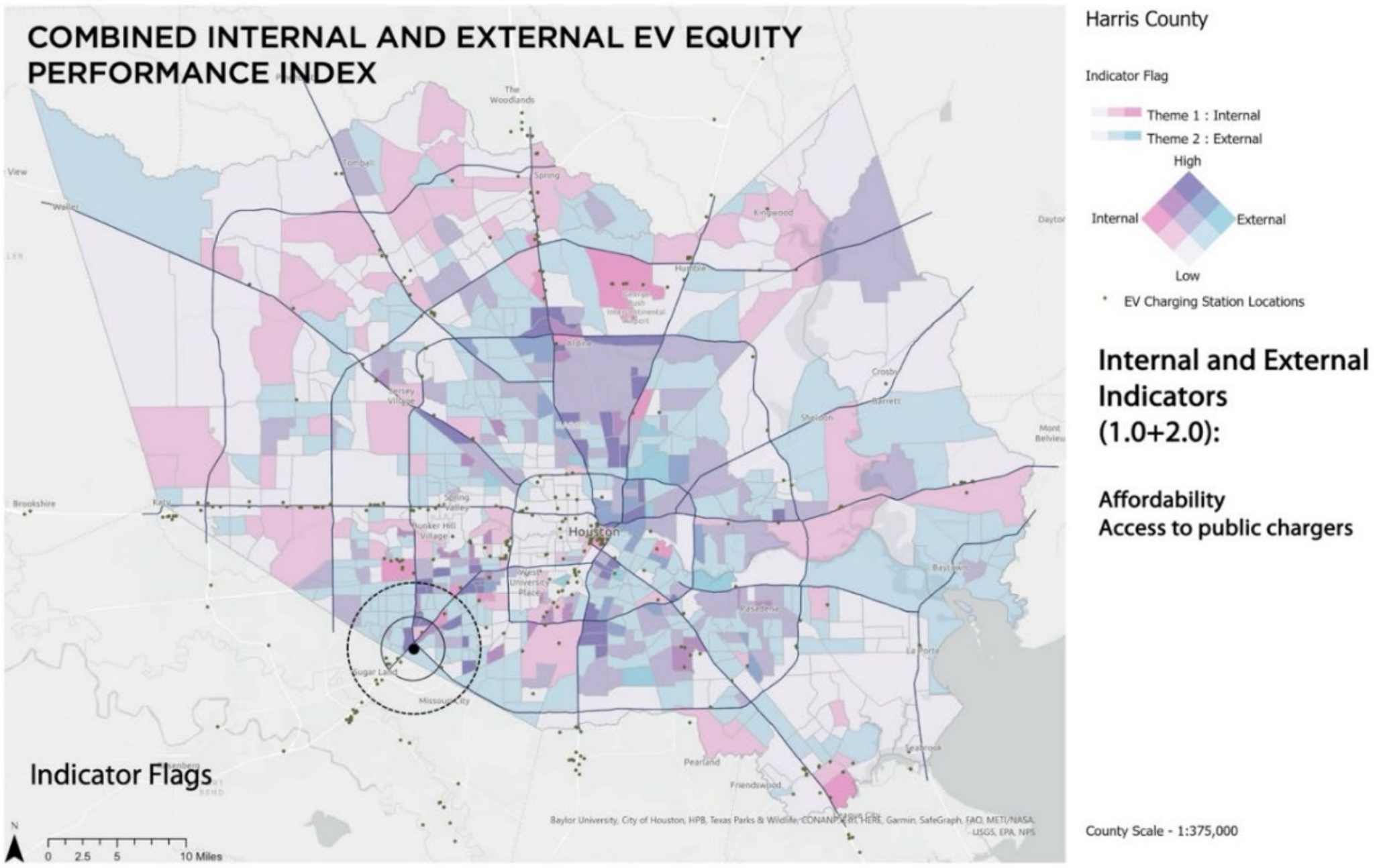

*Figure 1.* EV Equity Index.

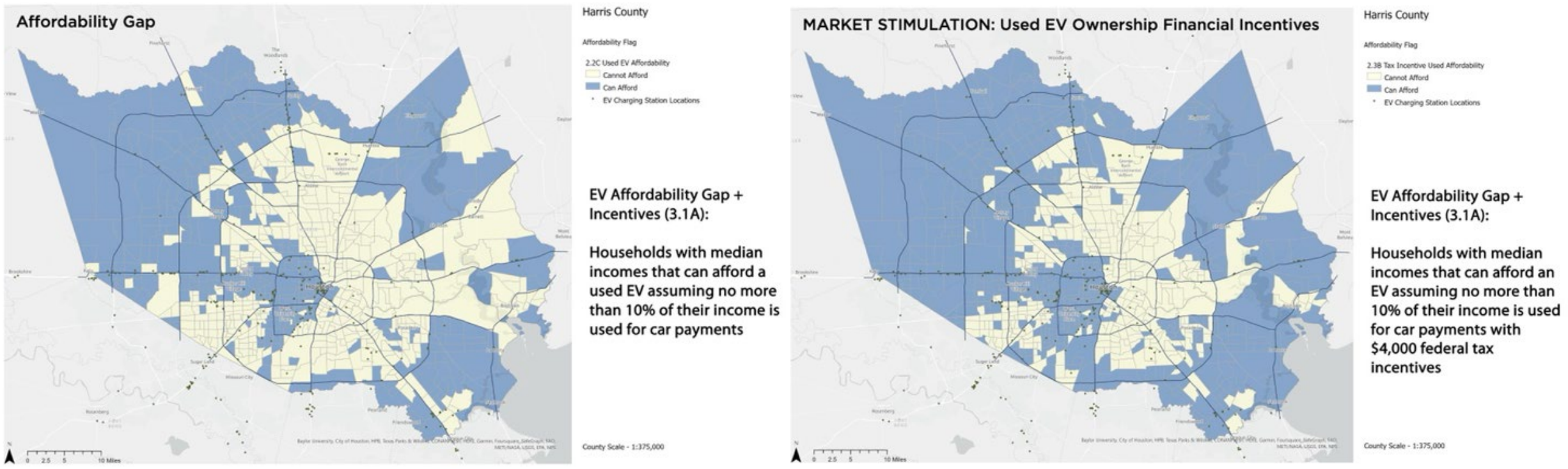

*Figure 2.* EV Ownership Affordability Gap.

3.3 EV ADOPTION CURVES AND CHARGING INFRASTRUCTURE

The Houston region has growing light vehicle EV fleet but is lagging in public access charging. In 2024, 10.2% of U.S. auto sales were EVs. The last quarter of 2024 was 10.9% (Alliance for Automotive Information, 2025). The Houston region EV adoption is growing faster than the national average. EV sales in the fourth quarter of 2024 were 14.53% the first half of 2024, and nearly 16% in the fourth quarter (Evolve Houston, 2024). In 2024, the national ratio of EV to (nonhome) public charger was approximately 1:22 (Price, 2025) compared to a 1:25.8 ratio in Texas (DOE, 2025).

## 4. Policy Options

In the Climate Action Planning process, the City of Houston explored policy options for meeting emission reduction targets. Variables included EV adoption rate assumptions, VMT reduction, and population growth. The H-CAP process included development of a business as usual (BAU) estimate and development of policy scenarios to inform community discussion, and a preferred scenario that became the basis for the H-CAP on-road transportation goals and strategies.

4.1 BUSINESS AS USUAL

A BAU estimate was prepared using data from the 2014 Baseline and 2050 population growth estimates from the Texas Development Board. The BAU assumed there was no policy or technology innovation and the VMT growth tracked population and fuel efficiencies stayed the same as the baseline.

4.2 POLICY SCENARIOS

Four policy scenarios were developed to explore the types of assumptions and actions would provide progress towards meeting a net zero GHG on-road transportation 2050 target. Scenario one assumed Houston would meet the Obama era car and small truck 2025 and truck 2027 standards. Scenario 2 assumed the same fuel efficiency standards but with an added 30% reduction in VMT. The third scenario assumed vehicles would be meeting “technical limits” of fuel efficiency and there would be a 40% reduction in VMT. The H-CAP Transportation Subcommittee discussion resulted in a preferred scenario 4 - - “technical limits” and a 20% reduction in VMT. This would reduce GHG from on-road transportation by 70% in 2050 (Figure 3).

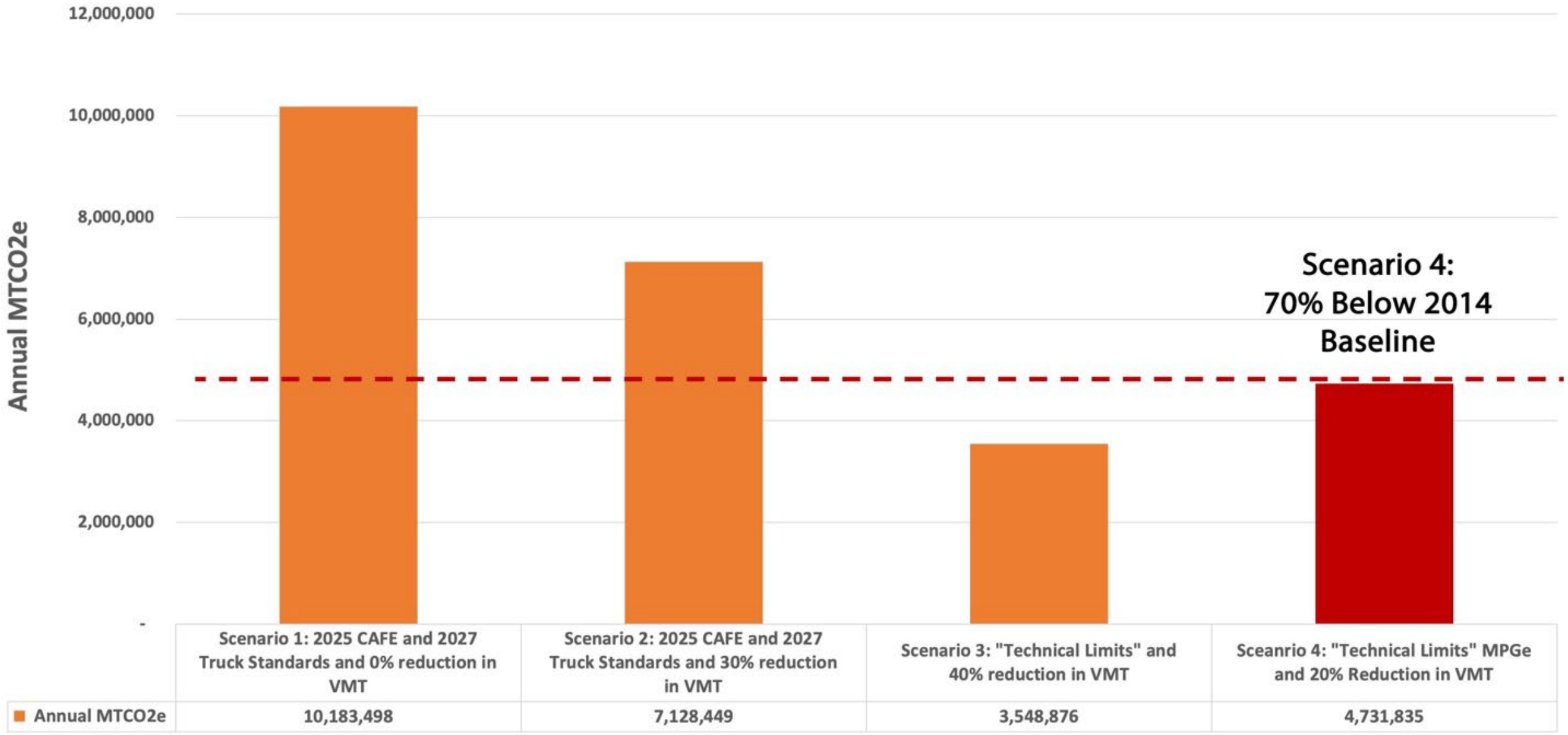

*Figure 3.* On-Road Transportation Policy Scenarios.

4.3 SUPPLY-SIDE IMPLICATIONS

Even with an assertive demand reduction policy for on-road transportation, there still is a performance gap. To close this gap, on-road transportation would need an estimated 15,490 gWh of renewable energy annually (Race, Houston-CAP On-Road Transportation GHG Emissions: Baseline, BAU, and Policy Scenario Analysis, 2019). Figure 4 indicates the 2030, 2040, and 2050 goals for GHG reductions and off-sets in renewable energy to meet 2050 net zero targets. Using data from National Renewable Energy Lab (NREL) for average gWh per acre, Houston's offsets would require about 67 square miles of community solar.

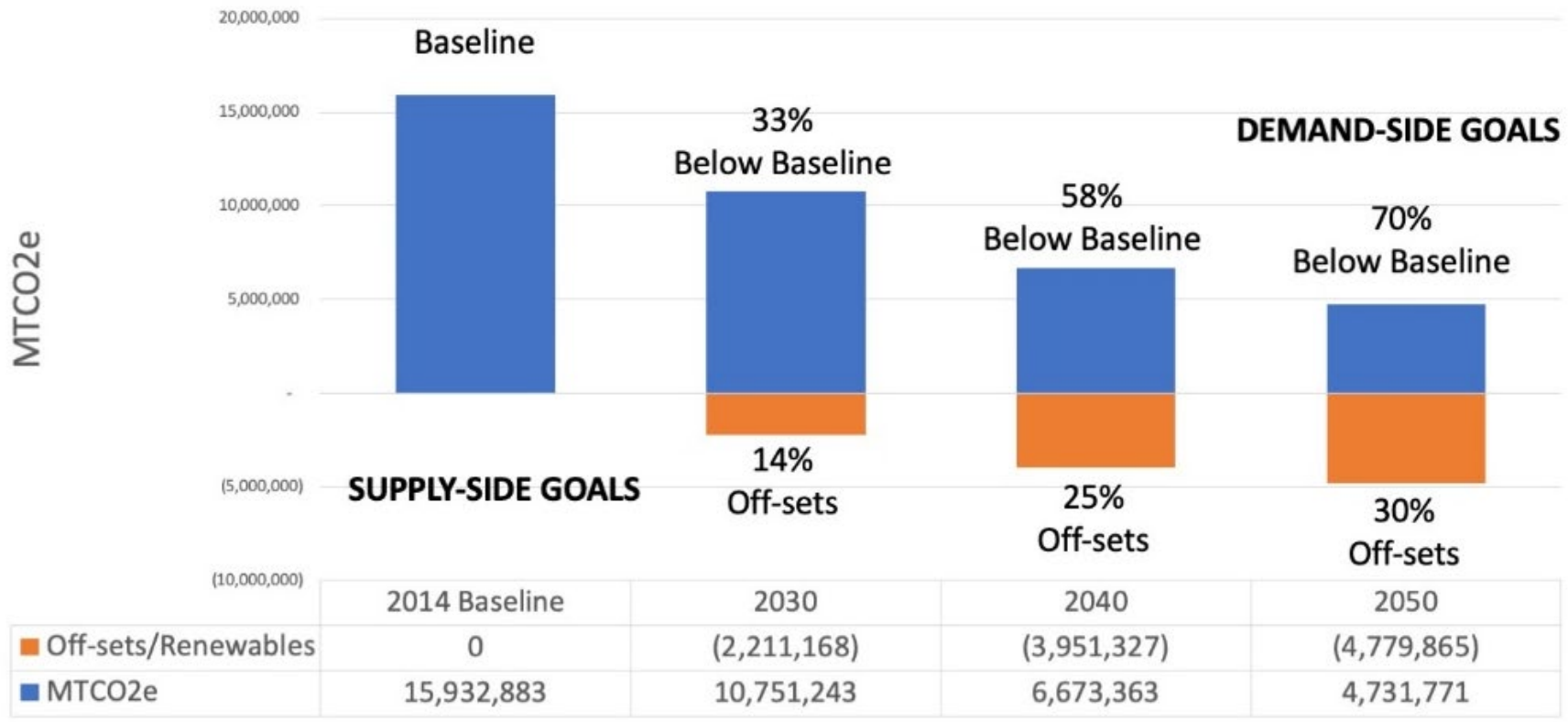

*Figure 4.* On-Road Transportation Goals.

## 5. Use of ITS and Smart Transit and Intermodal Optimization

The assumed 20% reduction in VMT is an aggressive goal. Sprawling Houston would have a lower VMT than Portland, Oregon. To achieve this, Houston will need to implement land use planning around transit, improve intermodal performance, and optimize public transit. As a low-density city, Houston must increase both transit penetration and its intermodal performance. METRO owns 30 park and ride sites. Many of these can be planned to support electrification of vehicles, diversify modes, and improve mode transfers (Figure 5).

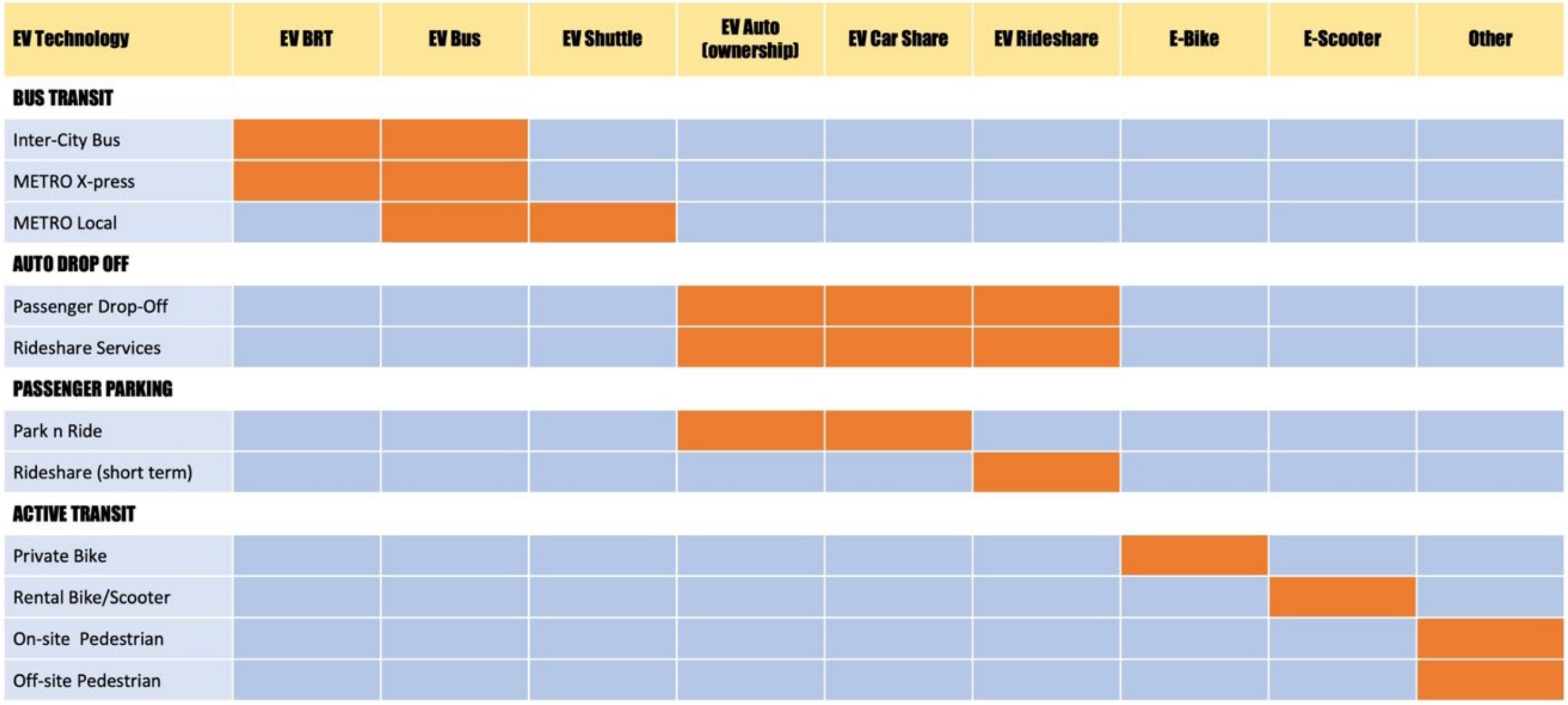

*Figure 5.* Intermodal Transportation Hubs.

### 5.1 INTELLIGENT TRANSPORTATION SYSTEM

Applied technologies could include expanded use of Intelligent Transportation Systems (ITS), optimization of streets and highways and improving access to intermodal transportation hubs. Both the Texas Department of Transportation (TxDOT) and City of Houston Department of Public Works (DPW) have extensive Bluetooth sensor networks. They have installed Bluetooth trackers along highways and thoroughfares. TxDOT has set up the TranStar sensor network to measure travel speed and congestion on Houston highways. The sensors track Bluetooth devices (phones) as an electronic address. Each roadside reader senses addresses emitted by enabled devices as they pass (Houston TranStar, 2025). DPW integrates ITS infrastructure to improve safety, efficiency, and mobility. The system includes Bluetooth devices, Ethernet switches and routers, fiber optic cables, Closed-Circuit Television (CCTV) cameras, mid-block counters, Dynamic Messaging Signs, and enhanced detection devices. ITS reduces delays at intersections, thereby controlling speeds and improving travel time (Houston Public Works, 2025).

### 5.2 SMART TRANSIT HUBS

METRO has 30 park and ride sites that can become smart transit hubs. Their design can reflect expected functions and approach to data management. A templet for smart transit hubs would provide common framework for planning, defining, and integrating intelligent transportation systems. This approach can serve as METRO’s roadmap to evolving into smart transit service provider. The hub could be organized around four services:

- Smart Charging Infrastructure–strategies for placing charging stations across the park and ride facilities to ensure convenient access for electric vehicle users.
- Smart Parking Solutions–leveraging real-time data and advanced parking management systems to optimize parking space availability.
- METRO Pass System–enabling park and ride users to seamlessly transition between various transportation modes (buses, trains, and parking) using a single payment method.
- Intermodal Connectivity–allowing commuters to switch between EVs, buses, or trains with minimal effort and ensuring a smooth and efficient commuting experience supported by digital displays and user-friendly mobile applications.

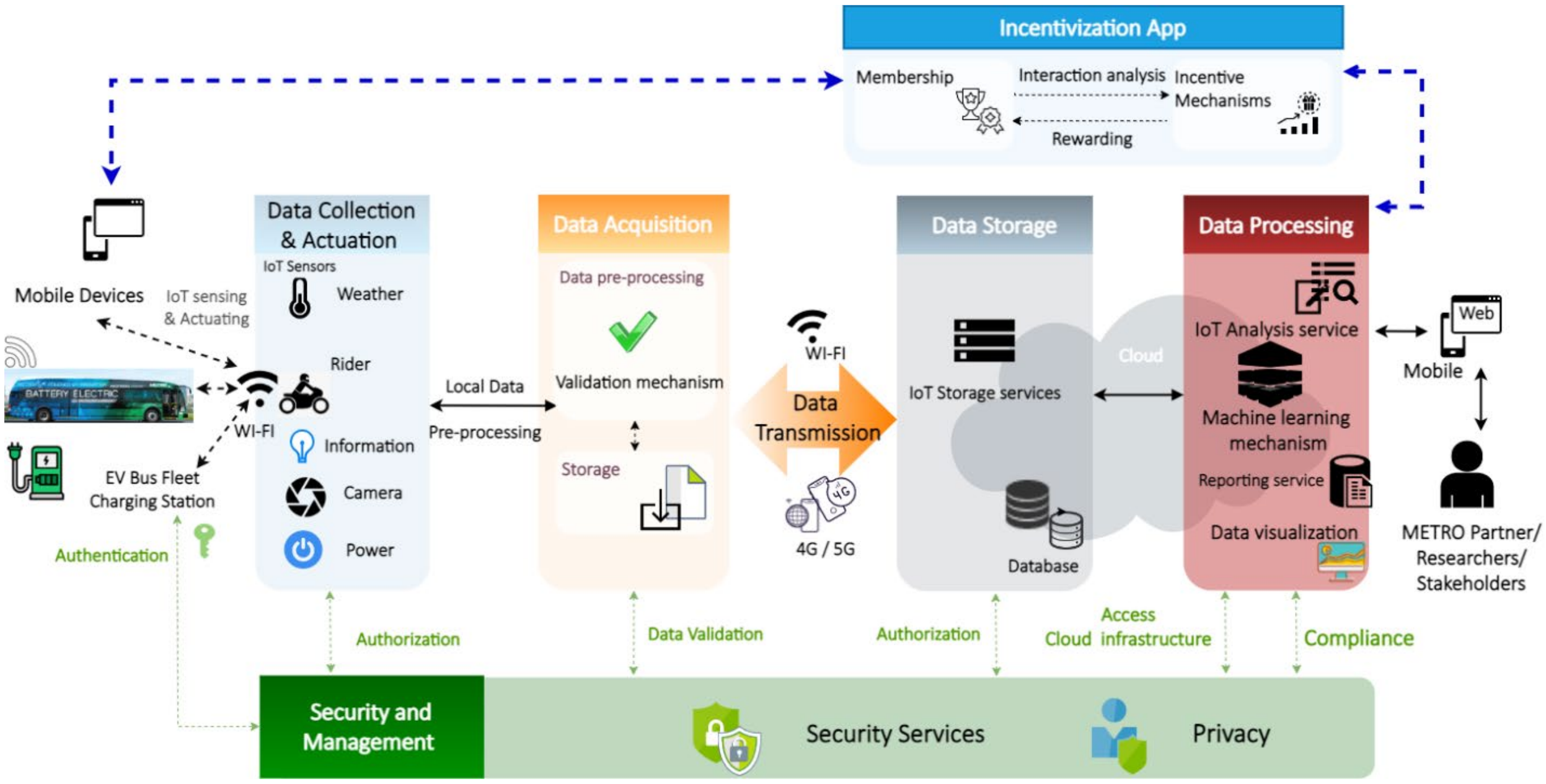

*Figure 6.* Smart Transit Hub Digital Services.

- The proposal applies smart technologies that improve the reliability, comfort, and affordability of transit services. The smart transportation intermodal hubs can be organized into four layers and supported with incentives and specialized security protocols (Figure 6).

- The Data Collection Layer enables the METRO system to collect data from its EV bus fleet and charging station and information about riders to improve their experience. Data will include images, GPS, Weather, etc. This layer involves privacy protection and authentication.
- The Data Acquisition Layer receives data from the data collection layer and stores it in the device and buses to apply the data validation mechanism to filter out unqualified data and to hide the private information in the images and videos before submitting it to the cloud. The data will be transmitted to the next layer after it is authorized. If we're using smartphone apps to collect data, users will be able to make decisions about what data is authorized and what modes of transportation is he interested in.
- The Data Storage Layer is a cloud server that consists of aggregating data that has been received from the Data Acquisition Layer.

- The Data Processing Layer is a cloud computing layer that provides resources to process aggregated data to verify its reliability and eliminate redundant data. This layer also applies data analysis, reporting, and data visualization services to retrieve as much knowledge as possible before providing it to the stakeholder application and designing incentives to engage riders to use public transport integrated with intermodal transport.
- Incentivization App presents the incentive mechanism service. Its main role consists of applying the game design elements we proposed for this architecture, acting as a link between the cloud server which provides incentives to users who favor EV-related services.
- Security Management is responsible for providing a secure service that combines varied security aspects such as data security in terms of data integrity, data access, authentication, authorization, and data privacy.

### 5.3 SMART EV CHARGING

The METRO's smart transit hubs will operate smart EV charging stations within the hub's Data Collection Layer, collecting usage patterns, occupancy data, energy consumption, and charging behavior (Oladimeji D, 2023). This integration will enable METRO managers to align charging services with real-time transit schedules and passenger flow. Using the data collected, the application processing layer will employ predictive analytics to manage grid load and use the learning experience to size the number of EV charging station around the METRO network.

Given that charging transactions and location data are exchanged through a network, the Security Management framework governs all aspects of authentication, encryption, and user consent. End-to-end data protection ensures that EV charging data can be safely leveraged for system optimization without compromising user privacy.

The data collected will enable the following smart functionalities:

- Implement Incentivization model that uses gamification by rewarding users who regularly charge at METRO hubs or synchronize their EV trips with public transport usage (Pilla, 2025). Through a unified METRO Pass System, users can view real-time charger availability, reserve slots, and pay using the same digital wallet used for parking or bus fare.
- Reduce range anxiety and support EV adoption. Many users who live in apartment and rely on public transportation system will find the charging station and good resource.
- Generate actionable insights for infrastructure planning and service improvement. This will strengthen the regional vision for sustainable, data-driven urban mobility (Michalakopoulou, 2025).

## 6. Policy Evaluation Framework Methods

Enabling technologies such as intelligent transportation systems, smart transit hubs, and EV charging infrastructure provide essential tools for advancing decarbonization and sustainable mobility. However, their true impact depends on careful evaluation and integration within broader policy strategies. To address this need, we propose a policy evaluation framework that incorporates socio-economic indicators, technology adoption patterns, and spatial mobility data into a unified assessment platform.

A central innovation of this framework is the use of a Unity 3D-based simulation environment. Unity 3D is particularly well-suited for this application, as it enables the development of interactive, high-fidelity simulations that accurately represent urban mobility dynamics. This capability allows decision-makers to explore, visualize, and evaluate transportation interventions in a safe and controlled virtual environment (Kassim, 2021) (Martinez, 2025). Ultimately, this tool enhances the rigor, transparency, and equity of policy design by enabling the testing of urban mobility strategies under realistic conditions.

### 6.1 ESTABLISHING A BASELINE GHG INVENTORY

A critical foundation of any rigorous policy evaluation framework is the establishment of a comprehensive baseline GHG emissions inventory for the transportation sector. This baseline serves not only as a diagnostic tool to quantify existing emissions, but also as a reference point for assessing the effectiveness of proposed interventions over time. To ensure both analytical accuracy and policy relevance, the inventory must capture emissions across multiple dimensions, including transportation modes, fuel types, spatial distribution, temporal variability, and socio-demographic characteristics. The process begins with the integration of high-resolution traffic activity data, typically derived from road segment-level vehicle counts, average travel speeds, and congestion patterns. These data are often sourced from transportation departments, GPS-based mobility datasets, or traffic sensor networks. Coupled with detailed vehicle registration records; disaggregated by vehicle type, fuel category, and model year. These inputs enable the estimation of vehicle fleet composition and usage patterns.

Origin-destination matrices, reconstructed from mobile phone GPS traces, transit smart card records, or travel surveys, offer insight into the spatial structure of daily travel demand. This information is particularly important for capturing the variation in travel intensity and modal split across different urban neighborhoods, income groups, and employment zones. The resulting inventory enables the construction of a temporally and spatially resolved emissions map that reflects both the current state of the transportation system and its embedded disparities. Such granularity is crucial for modeling the marginal effects of policy interventions—such as electrification, congestion pricing, or land use reform on specific subpopulations or regions. Furthermore, a robust baseline enhances transparency and accountability in long-term climate action planning, supporting both municipal compliance with emission reduction targets and equitable policy design.

### 6.2 TARGETS, GOALS, AND ACTIONS

A central component of the policy evaluation framework involves the articulation and operationalization of SMART (Specific, Measurable, Achievable, Relevant, and Time-bound) goals. These goals offer a structured and transparent approach for guiding transportation decarbonization efforts, ensuring accountability, and enabling iterative improvements over time. When clearly defined and contextually grounded, such goals help policymakers and stakeholders align investments and interventions with long-term climate, equity, and mobility objectives. A primary focus within this structure is the reduction of GHG emissions from the transportation sector. This aligns with federal and state-level carbon neutrality pledges and is supported by sector-specific carbon budgeting and lifecycle emissions modeling.

Achieving GHG reductions requires both technological transitions and behavioral shifts, including widespread electrification and mode substitution. Another critical dimension of this framework involves ZEV adoption goals. For example, a jurisdiction may set the goal of reaching 30% ZEV market share by 2035, with explicit provisions for equitable access across income brackets and geographic areas. Such targets are often embedded in state ZEV action plans and may be reinforced by incentives, infrastructure deployment mandates, and outreach campaigns targeting underserved populations. Income-based ZEV incentives and mobility credits can enhance participation from disadvantaged communities, mitigating existing disparities in EV adoption. In parallel, goals to reduce VMT address the underlying demand for car travel and aim to reduce overall transportation energy consumption. For example, Houston's goal is to reduce VMT by 20% reduction in VMT per capita by 2050. This is pursued through policies promoting transit-oriented development, active transportation infrastructure, and the integration of shared and public mobility options. Reductions in VMT not only cut emissions but also improve public health outcomes and reduce congestion. Complementary to these outcome goals are infrastructure and service provision benchmarks that support the adoption of cleaner and more efficient transport modes. These include adequate deployment of public EV charging stations. Such benchmarks serve as critical enablers of behavioral change by reducing range anxiety, improving transit reliability, and fostering mode shift. Each target and action are spatially and demographically contextualized to promote transportation justice. Geographic Information Systems (GIS) and socio-demographic data are used to map interventions to specific communities, accounting for factors such as income, racial composition, housing density, and transit dependency. This granularity ensures that policy interventions do not disproportionately benefit already advantaged populations and that the distributional impacts of decarbonization policies are explicitly assessed and addressed. By integrating these SMART goals within a broader equity-centered planning framework, cities can pursue transportation decarbonization in a way that is both technically effective and socially inclusive. The clear articulation of such goals enables continuous performance tracking, adaptive management, and public accountability essential components of any resilient and just climate action strategy.

### 6.3 UNITY 3D-BASED SIMULATION PLATFORM

A key innovation introduced in this study is the proposal to deploy a Unity 3D as a dynamic, extensible simulation platform to visualize and evaluate urban transportation scenarios (Smajic, 2020). Traditionally used in gaming and virtual reality, Unity 3D has emerged as a powerful tool for scientific modeling due to its real-time rendering capabilities, advanced physics engine, and support for interactive user interfaces (Weißmann, Edler, Keil, & Dickmann, 2023). By leveraging these strengths, the proposed framework bridges the gap between technical analysis and policy communication in transportation planning. Unity's simulation engine enables the real-time modeling of multimodal transportation systems embedded within realistic urban environments. Unlike conventional GIS-based simulations that often present static maps and graphs, Unity provides immersive 3D visualizations of city blocks, road networks, public transport systems, and vehicle movements (Figure 7). These simulations are grounded in agent-based modeling approaches, allowing for the representation of heterogeneous traveler behaviors, vehicle interactions, and infrastructure usage patterns. Such behavioral realism enhances the credibility of policy testing, particularly when assessing mode shift dynamics, congestion responses, and equity implications. The framework supports integration with geospatial data such as OpenStreetMap road networks, city zoning shapefiles, and high-resolution satellite imagery, ensuring spatial fidelity. Additionally, the system incorporates empirical datasets including GTFS transit schedules, travel demand models, synthetic population distributions, and vehicle emissions coefficients sourced from EPA's MOVES model. These inputs enable the construction of finely calibrated simulation scenarios that mirror real-world conditions.

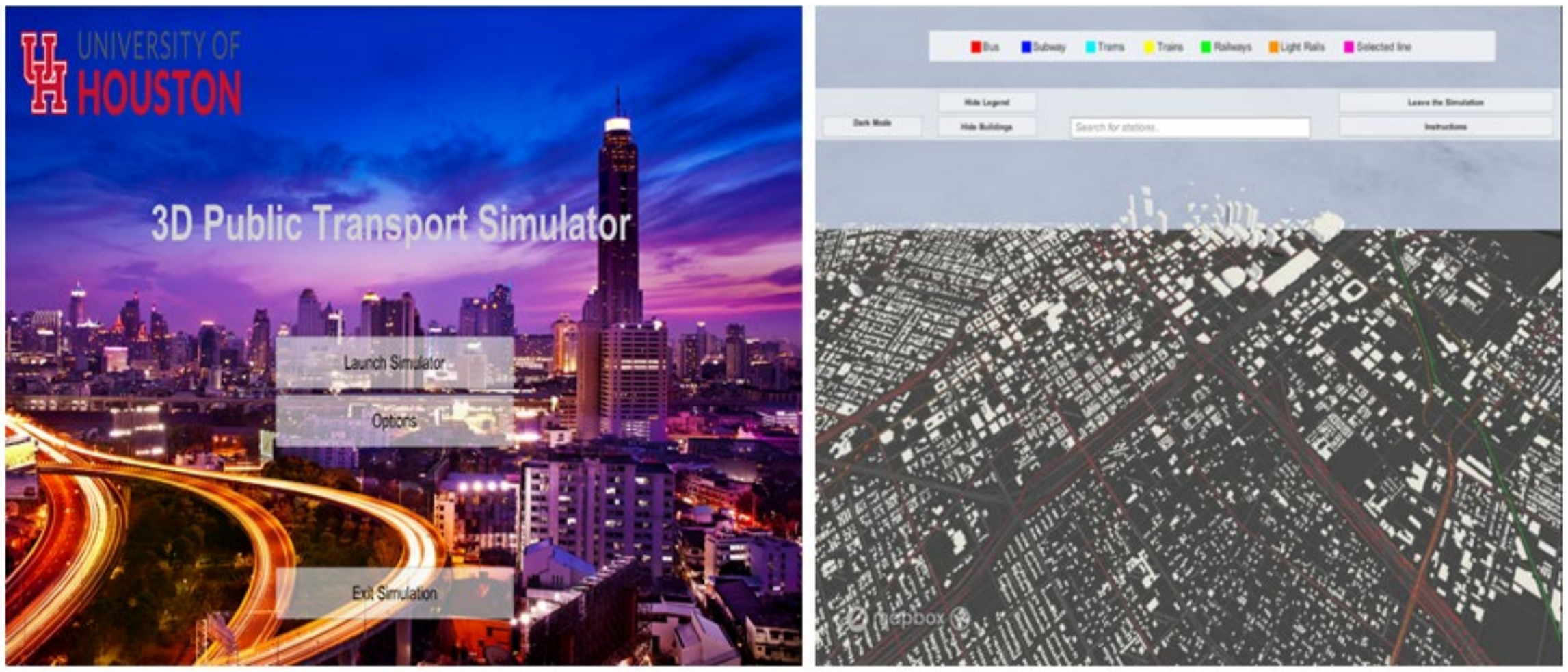

*Figure 7.* Public Transportation Simulator Unity platform example showing Houston as a case study to overlay traffic information on the top of 3D visualization.

A distinguishing feature of the Unity-based platform is its interactive feedback loop (Singh M, 2024) Policy levers such as congestion pricing, EV incentives, or curbside management rules can be modified in real time, with their impacts immediately visualized through animated changes in traffic density, mode shares, emissions levels, or charging network stress. This real-time responsiveness supports rapid prototyping and participatory engagement, making the platform suitable for stakeholder workshops and public consultations. Simulation modules developed within this framework allow for the exploration of a variety of policy and infrastructure interventions. For instance, low-emission zones can be simulated with varying levels of access restriction and enforcement rigor. Smart parking strategies can be analyzed in terms of search times and emissions trade-offs, while transit ridership changes can be tracked in response to ITS improvements or fare adjustments. EV charging behavior, including station choice and queuing delays, is modeled to assess grid load and range anxiety under different deployment scenarios. The versatility of Unity makes it applicable across a wide array of user groups. Urban planners can visualize the spatial implications of land-use policies and zoning changes; transportation agencies can evaluate system-wide impacts of ITS deployments or roadway redesigns; and policy makers can assess the effectiveness of fiscal and regulatory incentives. Community groups, especially those representing vulnerable populations, can better understand how specific scenarios influence local mobility patterns and environmental exposures. Unity's export functions enable the generation of emissions heatmaps, animated traffic flows, and dashboard interfaces tailored for both expert and non-expert audiences. These outputs enhance communication and transparency, supporting informed decision-making and participatory governance. Moreover, the scalability of the simulation engine from neighborhood-level modeling to city-wide or regional analysis makes it suitable for both pilot testing and comprehensive transportation planning.
Ultimately, Unity's incorporation into this analytical framework enhances the depth, clarity, and accessibility of comparative scenario evaluation. Its immersive and interactive capabilities complement traditional quantitative analyses, fostering broader stakeholder engagement and more resilient decision-making processes.

## 7. Conclusions

To equitably achieve 2050 greenhouse gas reduction targets from on-road transportation, the City of Houston will need to deploy strategies that utilize emerging technologies, switch fuels, and reduce vehicle miles traveled. Strategies that are needed to meet targets in Houston are transferable. Other large, low-density cities with high VMT and low transit mode splits can adopt these methods.

7.1 FOCUS ON WHAT IS IN PEOPLE'S DRIVEWAYS

In Houston, the 2014 Baseline emissions revealed that 89% of GHG emission from on-road transportation are parked in peoples' driveways (Figure 8). This includes gasoline and diesel fueled cars and small trucks as well as recreational vehicles and motorcycles. Successfully achieving the 2050 net target will require across the board EV adoption regardless of income. EV adoption rates look healthy in Houston, but the end of federal incentives in 2026 will reset market demand when many of Houston communities cannot afford auto loans or do not have access to credit. The City of Houston will need to continue to work with partners to maintain and expand EV ownership.

*Figure 8:*

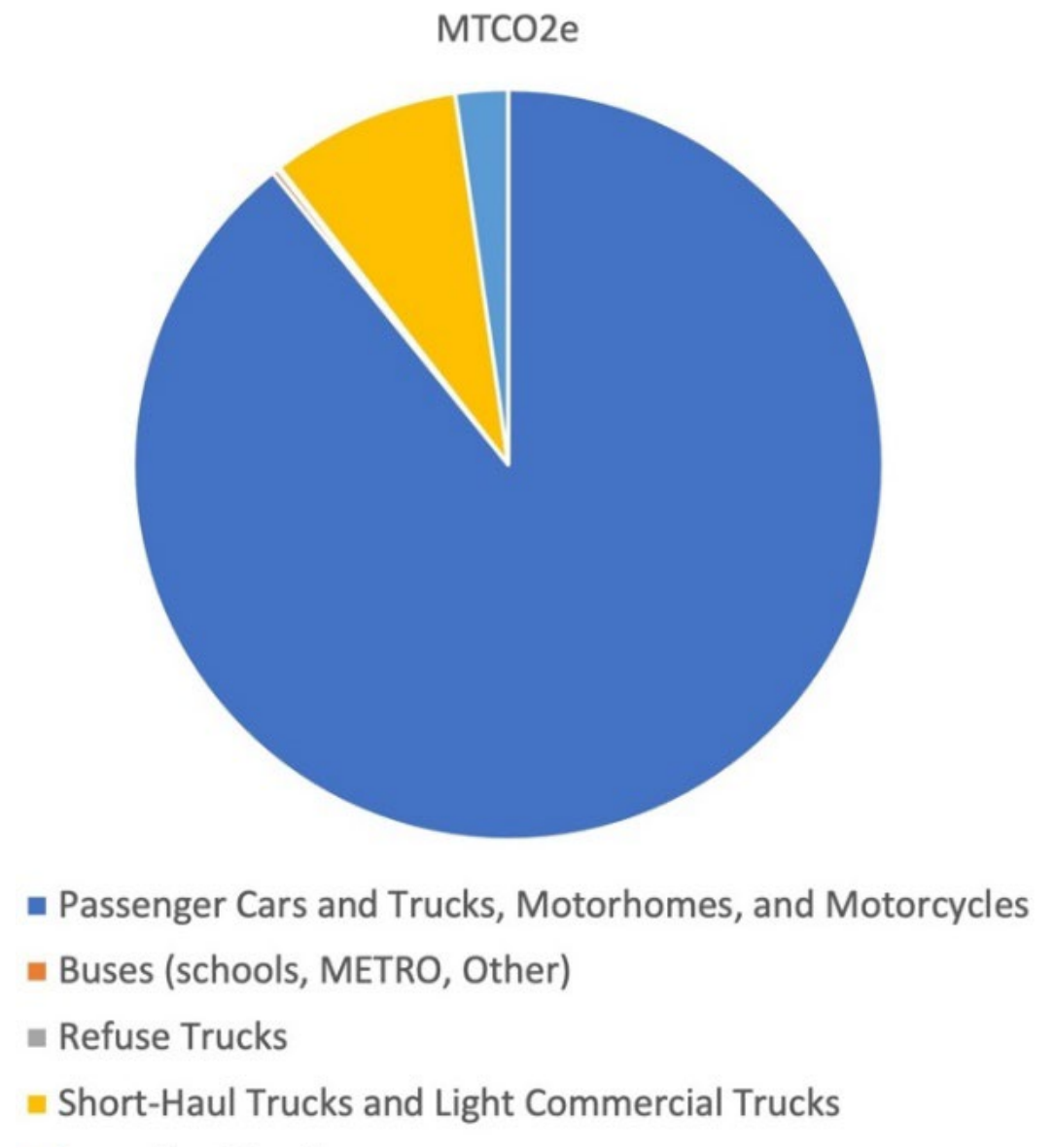

- Focus on passenger cars and trucks, and other personal vehicles (89 % of MTCO2e).
- Short-haul trucks and commercial small trucks are important (8%).
- Long-haul trucks are a supply chain issue -- national policy and regional partnerships are important (3%).
- Fleet vehicles --local policy (<1%)

*Figure 8*. Personal Transportation Largest Source of Transport GHG.

7.2 PROVIDE PUBLIC ACCESS AND APARTMENT CHARGING

Many of the next million people living in Houston will be living in apartments. Because 80-90% of charging happens at home, a major push for public charging will be necessary. In Houston, 57% households are renters and very few apartments have on-site charging. The Houston development code and accelerated permitting can incentivize apartment EV charging, while adding transit supporting densities. Public and private partners can create new business opportunities through public charging while increasing access for parts of Houston that are in charging deserts. A proactive approach is necessary.

7.3 IMPROVE METRO INTERMODAL EXPERIENCE

The H-CAP assumes a 20% reduction in VMT. This is a high bar for reductions in a sprawling city without a land use plan. Getting more people to use public transit will require improved first and last mile solutions, smoother intermodal experience, and financial incentives. METRO developing park and ride bus stations as transit and energy hubs can facilitate increased ridership, particularly as the region grows and infill development increases densities that makes Houston more walkable and social. The smart transit hub concept developed for METRO provides a roadmap to integrating smart parking, incentives, and more seamless travel.

7.4 MODEL STRATEGIES TO OPTIMIZE THE SYSTEM

The Simulator to Accelerate ZEV Adoption and Reduce VMT is a ready tool that Houston can use to optimize transportation services. The simulator can model routes and headway times to accommodate ridership growth, predict rider miles and cost of rider miles, measure emission of various routes, and optimize intermodal transportation hubs. This type of tool can be used to manage and plan the system to improve modal splits and reduce overall VMT.

7.5 SUPPLY SIDE SOLUTIONS

In addition to supporting fuel switching and VMT reductions, Houston and low-density cities will require increased capacity in the power grid and distribution system to provide for the added demand from the transportation sector. As the demand-side strategies will reduce the GHG emissions by and estimated 70%, there is still an off-set of 30% required. This will require an estimated equiveillance of 67 square miles of community solar - - or some combination of wind, distributed solar, and utility scale solar. State policies and regional grid operators need to plan for the added demand.

## Acknowledgements

This material is based upon work supported by the National Science Foundation under Grant NSF 21-535, Smart and Connected Communities program. Sponsor Award Number: 2126633.

## References

Alan Jenn, Katalin Springel, Anand R. Gopal, "Effectiveness of electric vehicle incentives in the United States,
Energy Policy," Volume 119, 2018, Pages 349-356,

Alliance for Automotive Information. (2025, March 28). Alliance for Automotive Innovation Reports New U.S. Electric Vehicle Data. Retrieved from Alliance for Automotive Information: https://www.autosinnovate.org/posts/press-release/2024-q4-get-connected-press-release

Bryan, L., & Landrigan, P. (2023). PM2.5 pollution in Texas: a geospatial analysis of health impact functions. Frontiers in Public Health, 11:1286755.

City of Houston. (2020). Climate Action Plan.

DOE. (2025, September 12). Alternative Fueling Station Counts by State. Retrieved from DOE: https://afdc.energy.gov/stations/states

Doust, M. (2018, November 29). More than 60 cities now use “gold-standard” global protocol to report on GHG emissions. Retrieved from C40 Cities: https://www.c40.org/news/more-than-60-cities-now-use-gold-standard-global-protocol-to-report-on-ghg-emissions/

Evolve Houston. (2024). RISE II: Houston‘s Regional Infrastructure Strategy for Electrification (RISE) Report. Houston.

Fawakherji, M., Benhaddou, D., Race, B., Carrollia, B., and Santamaria, W. (2023). Determinants of electric vehicle adoption and integration with public transportation through park-and-ride. International Wireless Communications and Mobile Computing (IWCMC)

Houston Public Works. (2025, September 16). Transportation Programs. Retrieved from Houston Public Works: https://www.houstonpublicworks.org/transportation-programs

Houston TranStar. (2025, September 16). Traffic Sensor Technologies. Retrieved from Houston TranStar: https://www.houstontranstar.org/faq/traffictech.aspx

icct. (2019, February 26). New study quantifies the global health impacts of vehicle exhaust. Retrieved from icct20: https://theicct.org/new-study-quantifies-the-global-health-impacts-of-vehicle-

exhaust/#:~:text=The%20study%20estimates%20that%20vehicle%20tailpipe%20emissions,worldwide%20in%202010%20and%20~385%2C000%20in%202015.&text=Exhaust%20from%20on%2Droad%20diesel%2

Kassim, M. a. (2021). The Design Of Augmented Reality Using Unity 3d Image Marker Detection For Smart Bus Transportation. International Journal of Interactive Mobile Technologies (iJIM), 33.

Li, C., & Managi, S. (2021). Contribution of on-road transportation to PM2.5. Nature, Sci Rep 11, 21320.

Maciag, M. (2013). Population Density for US Cities Statistics. Retrieved from Governing: https://www.governing.com/archive/population-density-land-area-cities-map.html

Martinez, V. M. (2025). Sustainable Intelligent Transportation Systems via Digital Twins: A Contextualized Survey. IEEE Open Journal of Intelligent Transportation Systems, 363-392.

Michalakopoulou, A. N. (2025). Decoding mobility hubs: Opportunities and risks underpinning their introduction for the contexts of transport and the wider society. Journal of Transport Geography, 104296.

Oladimeji D, G. K. (2023). Smart Transportation: An Overview of Technologies and Applications. Sensors (Basel), 8.

Pilla, P. T. (2025). Empirical evidence and policy insights from smart mobility hubs implementation for public office staff in Dublin City. Transportation Research Procedia, 83-90.

Price, L. (2025). U.S. light-duty vehicle charging infrastructure deployment through 2024. icct.

Race, B. (2019). Houston-CAP On-Road Transportation GHG Emissions: Baseline, BAU, and Policy Scenario Analysis.

Race, B. (2022, November 3). Equitable Access to EV-Mobility. Investigating the Challenges in Moving to US 2050 Climate Goals. Houston, TX: Texas Southern University.

Rodrigue, J.-P. (2024). Urban Density and Energy Consumption. New York: Routledge.

Singh M, K. J. (2024). Unity and ROS as a Digital and Communication Layer for Digital Twin Application: Case Study of Robotic Arm in a Smart Manufacturing Cell. Sensors (Basel), 17.

Smajic, A. (2020). Entwicklung und Erprobung eines interaktiven 3D - Stadtmodells am Beispiel des Personennahverkehrsnetzwerks der Stadt Frankfurt.

Taylor, D. E. (2014). Toxic Communities: Environmental Racism, Industrial Pollution, and Residential Mobility. New York University Press.

UN-Habitat. (2025, September 10). Urban Energy. Retrieved from UN-Habitat: https://unhabitat.org/topic/urban-energy

Weißmann, M., Edler, D., Keil, J., & Dickmann, F. (2023). Creating an Interactive Urban Traffic System for the Simulation of Different Traffic Scenarios. Appl. Sci, 13.